\documentclass[reprint, superscriptaddress, amsmath,amssymb, aps, prx, longbibliography,]{revtex4-1}
\usepackage{braket}
\usepackage{graphicx}
\usepackage{amsmath}
\usepackage{amssymb}
\usepackage{algorithm}
\usepackage{algpseudocode}
\usepackage{bbm}
\usepackage[colorlinks=true,urlcolor=blue,linkcolor=blue,citecolor=blue]{hyperref}
\usepackage[margin=0.75in]{geometry}

\newcommand{\tr}{\operatorname{tr}}

\begin{document}

\title{Capacity of Entanglement for Non-local Hamiltonian} 

\author{Divyansh Shrimali}
\affiliation{\(\)Quantum Information and Computation Group,\\Harish-Chandra Research Institute,
A CI of Homi Bhabha National Institute, Chhatnag Road, Jhunsi, Prayagraj  211019, India
}
\author{ Swapnil Bhowmick}
\affiliation{\(\)Quantum Information and Computation Group,\\Harish-Chandra Research Institute,
A CI of Homi Bhabha National Institute, Chhatnag Road, Jhunsi, Prayagraj  211019, India
}
 \author{Vivek Pandey} 
 \affiliation{\(\)Quantum Information and Computation Group,\\Harish-Chandra Research Institute,
A CI of Homi Bhabha National Institute, Chhatnag Road, Jhunsi, Prayagraj  211019, India
}
\author{ Arun Kumar Pati}
 \affiliation{\(\)Quantum Information and Computation Group,\\Harish-Chandra Research Institute,
A CI of Homi Bhabha National Institute, Chhatnag Road, Jhunsi, Prayagraj  211019, India
}
\date{\today}

\begin{abstract}
The notion of capacity of entanglement is the quantum information theoretic counterpart of the heat
capacity which is defined as the second cumulant of the entanglement spectrum. Given any bipartite pure state, we can define the capacity of entanglement as the variance of the modular Hamiltonian in the reduced state of any of the subsystems. Here, we study the dynamics
of this quantity under non-local Hamiltonian. Specifically, we address the question: Given an arbitrary non-local Hamiltonian 
what is the capacity of entanglement that the system can possess? As an useful application, we show that the quantum speed limit for creating the entanglement is not only governed by the fluctuation in the non-local Hamiltonian, but also depends inversely on the time average of square root of the capacity of entanglement. Furthermore, we discuss this quantity for a general self-inverse Hamiltonian and provide a bound on the rate of the capacity of entanglement. Towards the end, we generalise the capacity of entanglement for bipartite mixed states based on the relative entropy of entanglement and show that the above definition reduces to the capacity of entanglement for pure bipartite states. Our results can have several applications in diverse areas of physics.  
\end{abstract}

\maketitle

\section{Introduction}

Entanglement has potential applications
in quantum information science ranging from quantum computing, quantum communication and host of other areas such as condensed matter physics, high energy physics and even string theory\cite{Preskill2006,Jiang}. It is considered a very useful resource in information processing tasks. Over several years, how to create and quantify entanglement has been a subject of major explorations\cite{Horo,Sreetama2017}. Thanks to technological progress, now we can create entanglement between two or more number of particles in quantum optical systems~\cite{Optics-Boto}, ion traps \cite{Pedrozo}, superconducting systems \cite{tardigrade,20Q}, and NMR setups \cite{nmr}. How to create entanglement between more and more number of particles and distribute over long distances still continues to be quite challenging \cite{lrent}. Quantum entanglement between two particles can of course be created depending on the choice of the initial state and suitable non-local interaction between them. However, design of suitable interacting Hamiltonian is not always easy. This makes the production of entanglement a non-trivial task. Therefore, it is natural to ask the question, for a given non-local Hamiltonian, what is the best way of exploiting this Hamiltonian to create entanglement.
This was addressed in Ref.\cite{rate_ent} 

Entanglement entropy is quite a useful diagnostic tool which measures degree of quantum entanglement between subsystems in a many-body quantum systems \cite{qtm_ent}. A different quantity, called as the capacity of entanglement has been proposed to characterize topologically ordered states in the context of the Kitaev model \cite{capacity_def}. Given a pure bipartite entangled state \(\rho_{AB}\), the capacity of entanglement is defined as the second cumulant of the entanglement spectrum.
Thus, associated to a reduced density matrix, we can define the capacity of entanglement as the variance of the modular Hamiltonian in the mixed state. If $\{\lambda_i\}$'s are the eigenvalues of the reduced density matrix of one of the subsystem, than the entanglement entropy is defined as $S_{EE} =S(\rho_{A}) =-\tr(\rho_{A}\log{\rho_{A}}) = -\sum_{i}\lambda_i \log\lambda_i$. Now, the capacity of entanglement $C_{E}$ is defined as the second cumulant of this entanglement spectrum\cite{li}, i.e., the variance in the entanglement entropy operator. It is similar to the heat capacity of thermal systems and is given by \cite{cnxn_heat_cap,li,john} 
\[C_E = \sum_{i}\lambda_i \log^{2}\lambda_i - S^{2}_{EE}.\]
The above quantity can be thought of as the variance of the distribution of $\,-\log\lambda_{i}$ with probability $\lambda_i$, and thus it contains information about the width of the eigenvalue distribution of the reduced density matrix. We can gain insight on the whole spectrum by studying upto first two cumulants, i.e., the entanglement entropy and the capacity of entanglement. Defining a modular Hamiltonian as $K_A =-\log\rho_{A}$, they are the expectation value and the variance of $K_{A}$.
The capacity of entanglement has found useful applications in the conformal and the nonconformal quantum field theories \cite{Nandi,Verlinde}, as well as in models related with the gravitational phase transitions \cite{Kawabata,Yoshitaka,Verlinde,Arias,kohki}.

In this paper, we address the entanglement capacities 
for non-local Hamiltonians. To be specific, we answer the following
question: Given a non-local Hamiltonian, what is the capacity of entanglement for bipartite systems? 
We show that the entanglement rate is bounded by the fluctuation in the non-local Hamiltonian and the capacity of entanglement. In addition, the quantum speed limit for creating the entanglement depends inversely on the fluctuation in the non-local Hamiltonian as well as on the time average of the square root of the capacity of entanglement. Thus, the more the capacity of entanglement, the shorter the time duration system may take to produce the desired amount of entanglement. We illustrate the quantum speed limit for general two-qubit non-local Hamiltonian and find that our bound is indeed tight. Furthermore, we discuss the capacity of entanglement for self-inverse Hamiltonains and provide a bound on the rate of capacity of entanglement. Finally, we generalise the capacity of entanglement for bipartite mixed states based on the relative entropy of entanglement measure. This definition reduces to the capacity of entanglement for the pure bipartite states. This will open up its explorations for mixed states in future. We believe that our results can find applications in diverse areas of physics ranging from condensed matter systems to conformal field theories and alike.

The present paper is organised as follows. In Section II, we provide basic definitions and useful relations for the capacity of entanglement for pure bipartite states. In Section III, we discuss the capacity of entanglement for non-local Hamiltonians.   In Section IV, we prove that the entanglement rate is bounded by the capacity of entanglement and the speed of quantum evolution under the non-local Hamiltonian. We also provide a quantum speed limit for entanglement production or degradation and discuss how the capacity of entanglement helps in deciding the speed limit. In Section V, we discuss the capacity of entanglement for self-inverse Hamiltonians and provide a bound on the rate of the capacity of entanglement. In Section VI, we generalise the definition of the capacity of entanglement for bipartite  mixed states based on the notion of relative entropy of entanglement. Finally, in Section VII, we summarise our findings.

\section{Definitions and Relations}
Let \(\mathcal{H}\) represent a separable Hilbert space and $\dim(\mathcal{H})$  be the dimension of Hilbert space. Let us consider a bipartite  quantum system  described by state vector  $\ket{\Psi}_{AB}\in {\cal{H}}_{AB} = {\cal{H}}_{A}\otimes{\cal{H}}_{B}$ with unit norm. It is possible to express the state vector $\ket{\Psi}_{AB}$ as
\begin{equation}
   \ket{\Psi}_{AB}= \sum_{n}\sqrt{\lambda_{n}}\ket{\psi_{n}}_{A}\otimes\ket{\phi_{n}}_{B}, \label{equ:Schmidt_Decomposition}
\end{equation}  
where $\{\ket{\psi_{n}}\}_{A} $ and $\{\ket{\phi_{n}}\}_{B} $ are the Schmidt basis in ${\cal{H}}_{A}$ and ${\cal{H}}_{B}$, respectively and $\{\lambda_{n}\}$ are the non-negative real numbers with $\sum_{n}\lambda_{n}=1$. Eq.~\eqref{equ:Schmidt_Decomposition} is called the Schmidt decomposition of $\ket{\Psi}_{AB}$ and $\lambda_{n}$ are known as the Schmidt coefficients. If the Schmidt decomposition of $\ket{\Psi}_{AB}$ has  more than one non-zero Schmidt coefficients then we say that system $A$ and $B$ are ``entangled''. If there is only one non-zero Schmidt coefficient then the state is not entangled.

Let ${\cal B({\cal H}}_{AB} )$  denotes the algebra of linear operators acting on a finite–dimensional
Hilbert space ${\cal H}_{AB} $ of dimension ${\rm dim}({\cal{H}}_{AB})$ and let ${\cal D}({\cal H}_{AB} )$ denote the set of density operators for the bipartite system. The density operators are positive operators of unit trace acting on ${\cal H}_{AB}$.
For any state $\rho_{AB} \in {\cal D}({\cal H})$, if we can express $\rho_{AB}$ as $\rho_{AB} = \sum_{i} p_{i} {\rho_{i}}^{A} \otimes {\rho_{i}}^{B}$
then it is separable state, otherwise the mixed state is entangled one.
Given a density operator $\rho_{AB}$ associated with a bipartite quantum system $AB$, 
the reduced density matrix for subsystem $A$ (or $B$) is obtained by taking partial trace over subsystem $B$ (or $A$), i.e., $\rho_{A}= \tr_{B}(\rho_{AB})$. A physical quantity of system $A$ represented by a self-adjoint operator ${\cal{O}}_{A}$ on ${\cal{H}}_{A}$ is identified with a self-adjoint operator ${\cal{O}}_{A}\otimes{\cal{I}}_{B}$ on  ${\cal{H}}_{AB}$, where ${\cal{I}}_{B}$  is the identity operator on ${\cal{H}}_{B}$.  The expectation value of ${\cal{O}}_{A}\otimes{\cal{I}}_{B}$ on state $\rho_{AB}$ is given by $\tr(\rho_{A}{\cal{O}}_{A})$, where $\rho_{A}$ is the reduced density operator of system $A$.

The quantum relative entropy between two density operators  $\rho$ and $\sigma$ acting on the same Hilbert space $\cal{H}$ is defined as \cite{Umegaki1962}
 \begin{equation}
    S(\rho\Vert\sigma) := 
       \begin{cases}
                \tr(\rho(\ln{\rho}-\ln{\sigma})) &  \text{if}\  \operatorname{supp}(\rho)\subseteq \operatorname{supp}(\sigma), \\ 
                +\infty & \text{otherwise},
       \end{cases}
 \end{equation}
where $\operatorname{supp}(\rho)$ and $\operatorname{supp}(\sigma)$ are the supports of $\rho$ and $\sigma$, respectively.
The quantum relative entropy satisfies important properties: (i) $ S(\rho\Vert\sigma) \ge 0$ and  $S(\rho\Vert\sigma) =0$ iff $\rho= \sigma$, (ii) $\sum_i p_i  S(\rho_i \Vert \sigma_i ) \ge  S(\sum_i p_i \rho_i \Vert \sum_i p_i \sigma_i) $ and (iii) $ S(\rho\Vert\sigma)  \ge 
 S({\cal E}(\rho) \Vert {\cal E}(\sigma) ) $  for any completely positive trace preserving map ${\cal E}$.

Let us consider a composite system $AB$ with pure state $|\Psi\rangle_{AB}$. The amount of entanglement between subsystems $A$ and $B$ can be quantified via the entanglement entropy which is defined as the von Neumann entropy of the reduced density operator $\rho_{A}= \sum_{n}\lambda_{n}\ket{\psi_{n}}_{A}\bra{\psi_{n}}  $ (or $\rho_{B}$), i.e., 
\begin{equation}
 S_{EE} = S(\rho_{A})=  - \tr(\rho_{A} \log\rho_{A}) = -\sum_{n}\lambda_{n}\log\lambda_{n}
\end{equation}
which is invariant under local unitary transformations on $\rho_{A}$. 
The von Neumann entropy vanishes when density operator $\rho_{A}$ is a pure state. For a completely mixed density operator, the von Neumann entropy attains its maximum value of $\rm \log d_{A}$, where $\rm d_{A} = dim(\cal{H}_{A})$.

For any density operator $ \rho_{A}$ associated with quantum system $A$, we can define a formal “Hamiltonian” $K_{A}$, called the modular Hamiltonian, with respect to which the density operator $\rho_{A}$ is a Gibbs like state (with $\beta=1$) 
$$\rho_{A}=\frac{e^{-K_{A}}}{Z},$$ 
where $Z=\tr(e^{-K_{A}}).$ Note that any density matrix can be written in this form for some choice of Hermitian operator $K_{A}$. With slight adjustments in the above equation, the modular Hamiltonian $K_{A}$ can be written as $K_{A}=-\log\rho_{A}$.  In this case, the entanglement entropy of the system is equivalent to the thermodynamic entropy of a system described by Hamiltonian $ K_{A} $ (with $\beta=1$). 
Writing in terms of modular Hamiltonian $K_A =-\log\rho_{A}$, the entanglement entropy
becomes the expectation value of the modular Hamiltonian
\begin{equation}
    S_{EE} = -\tr(\rho_{A}\log\rho_{A}) = \tr(\rho_{A} K_{A})= \langle K_{A}\rangle \,.
\end{equation}

The capacity of entanglement is another information-theoretic quantity that has gained some interest recently\cite{Pawel,capacity_def}. It is defined as the variance of the modular Hamiltonian $K_{A}$\cite{capacity_def} in the state $\ket{\Psi}_{AB}$ and can be expressed as
\begin{align}
 C_{E}(\rho_{A}) & = \Delta K_{A}^2 =\bra{\Psi}(K_{A}\otimes \mathcal{I}_{B})^{2} \ket{\Psi}-\bra{\Psi}(K_{A}\otimes \mathcal{I}_{B})\ket{\Psi}^{2} \nonumber \\
 & = \tr[\rho_{A}(-\log\rho_{A})^2]-[\tr(-\rho_{A}\,\log\rho_{A})]^{2} \\
 & = \tr[\rho_{A}K_{A}^2]-[\tr(\rho_{A}K_{A})]^{2}\nonumber \\      \label{equ:Entanglement_Capacity}
 &=\langle K_{A}^{2}\rangle-\langle K_{A}\rangle^{2} = \Delta K_{A}^2.
\end{align}
The capacity of entanglement can also be defined in terms of the variance of the relative surprisal between two density matrices $V(\rho||\sigma)$\cite{boes2020variance}:  
\begin{equation}
    V(\rho||\sigma) = \tr(\rho\left(\log\rho-\log\sigma\right)^2)-\left(D(\rho||\sigma)\right)^2.
\end{equation}
If one of the density matrices becomes maximally mixed (i.e., either $\rho$ or $\sigma$ becomes I/d), then the
variance of the relative surprisal becomes the capacity of entanglement.

As shown in Ref.\cite{pati_sum},  uncertainty for any observable is a convex function. Given two or more Hermitian operators such as $O_1$ and $O_2$, the standard deviation or the uncertainty for observables satisfy $\Delta (p_1 O_1 + p_2 O_2) \le p_1 \Delta O_1 + p_2 \Delta O_2 $ for $0 \le p_i \le 1 \,(i=1,2)$ with $\Delta O_{i}=\sqrt{\langle O_{i}^{2} \rangle-\langle O_{i}\rangle^{2}}$. This shows that adding two or more observables always reduces the uncertainty. If we define the standard deviation in the modular Hamiltonian as uncertainty in the entanglement operator, then for any two modular Hamiltonian $K_1$ and $K_2$, we will have
\begin{equation}
\Delta(\sum_{i}p_i K_i)\leq\sum_{i}p_{i}\Delta K_{i},   
\end{equation}
where $K_i = -\ln{\rho_i }$. This property has an interesting implication when we have a modular Hamiltonian undergoing some variation. Suppose, we allow a variation in the modular Hamiltonian as $K \rightarrow K'= K + x V$, where $V$ is the additional term in the modular Hamiltonian and $x$ is a real parameter. Then, the following relation holds true, i.e., $\Delta K' \le \Delta K + x \Delta V$.  

For the sake of completeness, we mention the following properties  which are applicable for \(C_{E}\) on account of having similar form as the relative surprisal between two density matrices:
\begin{enumerate}
    \item Additivity under tensor product:
    \[C_{E}(\rho_{A}\otimes\rho_{B}) = C_{E}(\rho_{A})+C_{E}(\rho_{B}).\]
    \item Positivity : \(C_{E}(\rho) \geq 0\).
    \item Uniform Continuity:
    \[|C_{E}(\rho)-C_{E}(\rho^{'})|^{2}\leq\xi\log^{2}{d}\cdot D(\rho,\rho^{'})\] for $\xi$ some constant and $l_1$ trace norm  between states $D(\rho,\rho^{'})$.
    \item $C_{E}(\rho) = 0$ if and only if all non-zero eigenvalues of $\rho$ are the same. Such states are termed as flat states. Examples include any pure state or maximally mixed state.
    \item Corrections to subadditivity:
    \[C_{E}(\rho)\leq C_{E}(\rho_1)+C_{E}(\rho_2)+\chi\log^{2}{d}\cdot f(I_{\rho})\] for any bipartite state $\rho$ with marginal states $\rho_1$, $\rho_2$ and mutual information $I_{\rho}$, with constant $\chi$ and \(f(x)=\max(x^{1/4},x^{2})\).
    \item For fixed dimensions $d\geq 2$, the state $\rho_d$ with maximal variance has the spectrum
    \[spec(\rho_{d}) = \Big( 1-r,\frac{r}{d-1},\cdots,\frac{r}{d-1} \Big)\]
    with $r$ being the unique solution to \[(1-2r)\ln\big(\frac{1-r}{r}(d-1)\big)=2\] We get \(\frac{1}{4}\log^{2}(d-1)< C_{E}(\rho_{d})<\frac{1}{4}\log^{2}(d-1)+\frac{1}{\ln^{2}(2)}\), and for the limit of large $d$, $r\approx\frac{1}{2}$.
\end{enumerate}
For further details and proofs regarding the above properties, readers are advised to go through Ref.~\cite{boes2020variance}.

\section{Capacity of Entanglement for NonLocal Hamiltonians}
The dynamics of entanglement under two-qubit nonlocal Hamiltonian has been adressed in Ref.\cite{rate_ent}. In this section, we address the following question: What is the capacity of entanglement for arbitrary two-qubit non-local Hamiltonian? Further, we also discuss about the rate of the capacity of entanglement for the non-local Hamiltonian. For any two-qubit system, the non-local Hamiltonian can be expressed as (except for trivial constants)
\begin{equation}
    H = \vec{\alpha}\cdot\vec{\sigma}^{A}\otimes\mathcal{I}_{B}+\mathcal{I}_{A}\otimes\vec{\beta}\cdot\vec{\sigma}^{B}+\sum_{i,j=1}^{3} \gamma_{ij}\sigma_{i}^{A}\otimes\sigma_{j}^{B}, \label{bipartite_Hamiltonian} 
\end{equation}
where $\vec{\alpha},\vec{\beta}$ are real vectors, $\gamma$ is a real matrix and, $\mathcal{I}_{A}$ and $\mathcal{I}_{B}$ are identity operator acting on $\mathcal{H}_{A}$ and $\mathcal{H}_{B }$. The above Hamiltonian can be rewritten in one of the two standard forms under the action of local unitaries acting on each qubits \cite{local_unitary,rate_ent}. This is given by
\begin{equation}
H^{\pm} = \mu_1\sigma_1^{A}\otimes\sigma_1^{B}\pm\mu_2\sigma_2^{A}\otimes\sigma_2^{B}+\mu_3\sigma_3^{A}\otimes\sigma_3^{B},
\label{Non_loc_hamiltonian}
\end{equation}
where \(\mu_1\geq\mu_2\geq\mu_3\geq 0\) are the singular values of matrix $\gamma$\cite{rate_ent}. Using the Schmidt-decomposition, 
any two qubit pure state can be written as
\begin{equation}
    \ket{\Psi}_{AB} = \sqrt{p} \ket{\phi} \ket{\chi} + \sqrt{1-p} \ket{ \phi^{\perp}} \ket{\chi^{\perp}}.
    \label{2qubit_general}
\end{equation}
We can utilize the form of Hamiltonian in Eq.~\eqref{Non_loc_hamiltonian} and choosing $H^{+}$ (i.e. assuming $\det(\gamma)\geq 0$) to evolve the state in Eq.~\eqref{2qubit_general} without loosing any generality \cite{rate_ent}. To further showcase a specific example, let us choose \(\ket{\phi}=\ket{0}\) and \(\ket{\chi}=\ket{0}\). Thus, the state at time $t=0$ takes the form
\begin{equation}
 \ket{\Psi(0)}_{AB}=\sqrt{p}\ket{0}\ket{0}+\sqrt{1-p}\ket{1}\ket{1}.  
 \label{equ:time zero state}
\end{equation}
Under the action of the non-local Hamiltonian, the joint state at time $t$ can be written as ($\hbar=1$)
\begin{align}
\ket{\Psi(t)}_{AB} = e^{-iHt}\ket{\Psi}_{AB} =\alpha(t)\ket{0}\ket{0}+\beta(t)\ket{1}\ket{1},
\label{equ:time evolved state}
\end{align}
where $\alpha(t) = e^{-i \mu _3 t} \left(\sqrt{p} \cos (\theta t) -i \sqrt{1-p} \sin( \theta t) \right)$ , $\beta(t) =e^{-i \mu _3 t} \left(\sqrt{1-p} \cos(\theta t) -i \sqrt{p} \sin(\theta t) \right)$ and $\theta = (\mu _1-\mu _2)$.
To evaluate the capacity of entanglement, we would require the reduced density matrix of the two qubit evolved state, $\rho_{A}(t)=\tr_{B}(\rho_{AB}(t))$, which is given by
\begin{align}
   \rho_{A}(t) &=  \lambda_{1}(t) \ket{0}\bra{0} + \lambda_{2}(t) \ket{1}\bra{1}, 
\end{align}
where $\lambda_1(t) = \left|\alpha(t)^2\right| $ and $\lambda_2(t) = \left|\beta(t)^2\right|$ with
\begin{align}
    \lambda_{1}(t) &= \frac{1}{2} \left[1 - (1-2 p) \cos \left(2\theta t\right) \right], \nonumber\\
    \lambda_{2}(t) &= \frac{1}{2} \left[1 + (1-2 p) \cos \left(2\theta t\right) \right]. \nonumber
\end{align}
The capacity of entanglement at a later time t can be calculated from the variance of modular Hamiltonian $ K_{A}$. This is given by
\begin{align}
  C_{E}(t)&=\tr(\rho_{A}(t)(-\log\rho_{A}(t))^2)-(\tr(-\rho_{A}(t)\log\rho_{A}(t)))^{2}\, , \nonumber\\
  &  = \sum_{i=1}^{2}\lambda_{i}(t) \log^{2}\lambda_{i}(t) - \left(-\sum_{i=1}^{2}\lambda_{i}(t) \log\lambda_{i}(t)\right)^{2} .
\end{align}

\begin{figure}[h!]
    \centering
    \includegraphics[width=9cm]{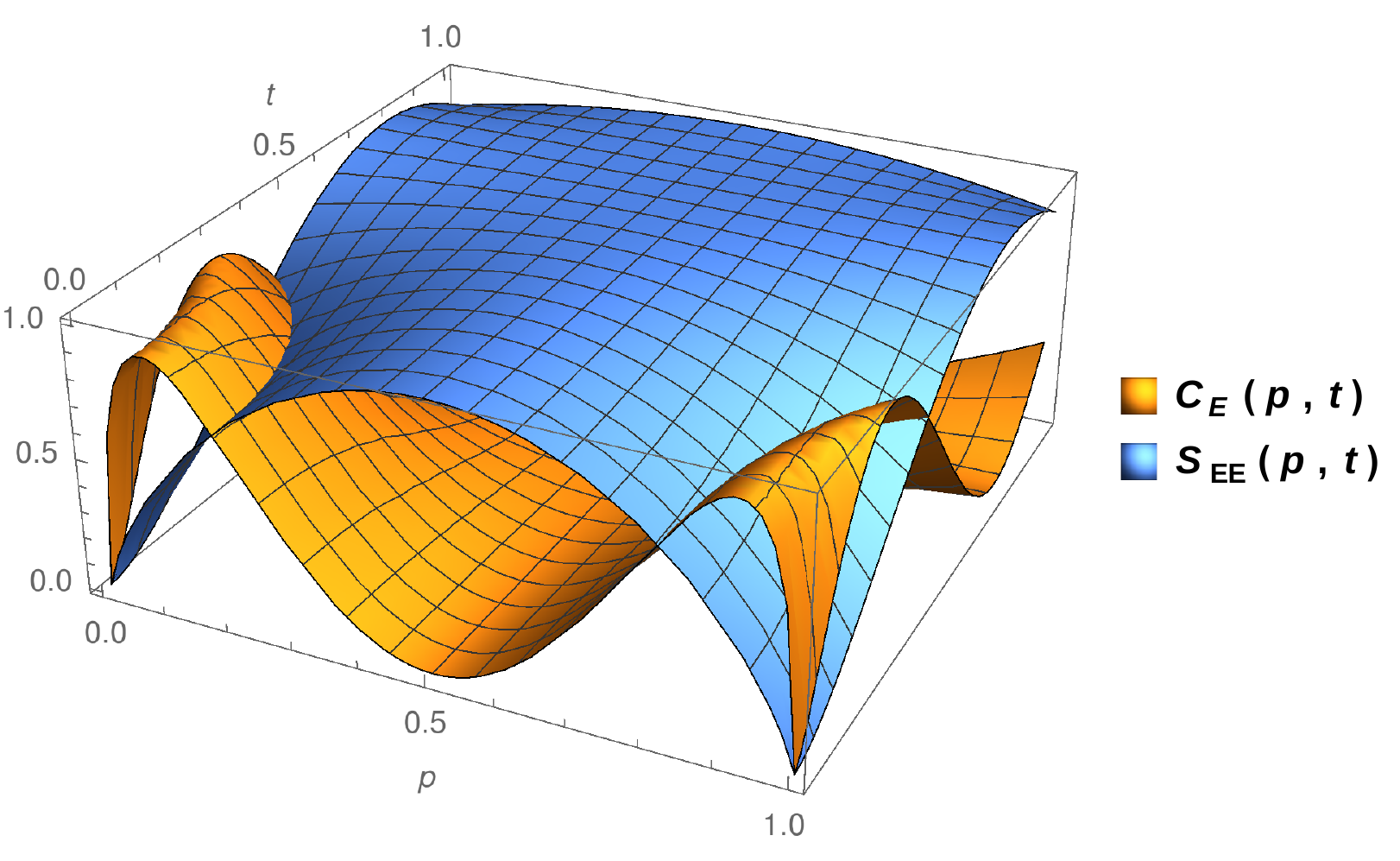}
    \caption{Plot for capacity of entanglement($C_E $) and entanglement entropy ($S_{EE}$) vs $p$ and $t$ taking \(\theta =1\).}
    \label{fig:C_E_S_p_t}
\end{figure}
 
In order to quantify the entanglement production, we can define the entanglement rate $\Gamma$ as defined in Ref.\cite{rate_ent}, i.e.,
\begin{equation}
    \Gamma(t)=\frac{{\rm d} S_{EE}(t)}{{\rm d} t}=\frac{{\rm d} S_{EE}(t)}{{\rm d} p}\frac{{\rm d} p}{{\rm d} t}.
    \label{rate_capacity}
\end{equation}
The assertion is that this quantity depends upon the entanglement $S_{EE}$ which depends upon some parameter $p$ and the rate of the Schmidt coefficient. The condition(s) to obtain a maximal entanglement rate are of interest for which two things are of significance. First, for a given value of $S_{EE}$ of two-qubit system, we find $|\Psi_{E}\rangle$, for which the interaction produces maximum rate $\Gamma_{E}$ and, the maximal achievable entanglement rate $\Gamma_{max}=\max_{E}\Gamma_{E}$ with corresponding state $|\Psi_{max}\rangle$.

Let us evaluate objects defined above for an arbitrary Hamiltonian \(H\). Using the Schmidt-decomposition of the state $|\Psi(t)\rangle$
\begin{equation}
    \ket{\Psi}_{AB} = \sqrt{p}|\phi\rangle |\chi\rangle + \sqrt{1-p}|\phi^{\perp}\rangle |\chi^{\perp}\rangle\, ,
    \label{Schmidt_bipartite}
\end{equation}
where $\langle\phi|\phi^{\perp}\rangle=0=\langle\chi|\chi^{\perp}\rangle $ and $p\leq\frac{1}{2}$. The entanglement measure $S_{EE}$, must depend only on the Schmidt coefficient $p$, given the fact that it must be invariant under local unitary operations. If we choose entropy of entanglement as $S_{EE}$, the entropy of reduced density operator of one of the qubits is given by
\begin{equation}
    S_{EE}(p)=-p\,\log_2 (p) - (1-p)\,\log_{2}(1-p).
\end{equation}
Operationally, $S_{EE}$ quantifies the amount of EPR entanglement contained asymptotically in a pure state $|\Psi\rangle_{AB}$, thus $S_{EE}$ gives a ratio of maximally entangled EPR state $|\Psi^{-}\rangle_{AB} = \frac{1}{\sqrt{2}}(\ket{0}\ket{0}-\ket{1}\ket{1})$ which can be distilled from $|\Psi\rangle_{AB}$.

Considering the infinitesimal time evolution of Schmidt coefficient of two qubit state, we get
\[|\Psi(t+\delta t)\rangle=e^{i H\delta t}|\Psi(t)\rangle\,\approx\, (1-i H\delta t)|\Psi(t)\rangle.\]
The time evolution of the reduced state for the subsystem \(A\) is given by
\begin{equation}
    \rho_{A}(t+\delta t)=\rho_{A}(t)-i \delta t\, \tr_{B}([H,|\Psi(t)\rangle\langle\Psi(t)|]).
\end{equation}
Starting from $\rho_{A}|\phi\rangle=p|\phi\rangle$, then using the Schr\"{o}dinger equation, we have
\begin{equation}
    \frac{{\rm d}p}{\rm dt}=2\sqrt{p(1-p)}\,{\rm Im}(\langle\phi,\chi|H|\phi^{\perp},\chi^{\perp}\rangle).
\end{equation}
As $\Gamma$ is to be maximized, we can choose
\[\Gamma = f(p)|h(H,\phi,\chi)|,\] where
\[f(p)=2\sqrt{p(1-p)}S_{EE}^{'}(p)\,\text{and}\,\,h(H,\phi,\chi)=\langle\phi,\chi|H|\phi^{\perp},\chi^{\perp}\rangle.\]
Note that fixing $S_{EE}$, means fixing $p$ and so the maximum entropy corresponds to a state with some fixed $|\phi\rangle$ and $|\chi\rangle$. For any value $S_{EE}$ of entanglement, the state $|\phi\rangle$ and $|\chi\rangle$ for which maximum entanglement rate $\Gamma_E$ is obtained does not depend on $S_{EE}$, but only on the form of Hamiltonian \(H\).\\
Let $h_{max}$ be the maximum value of $|h|$, i.e.,
\begin{equation}
    h_{max}=\max_{||\phi||,||\chi||=1}\big|\langle\phi,\chi|H|\phi^{\perp},\chi^{\perp}\rangle\big|.
\end{equation}
Now, we need to drive the two qubit state with local operators so that for all time the corresponding state is the one with maximum rate  and we would then know how capacity of entanglement evolves with time.\\

Evaluating the capacity for entanglement for general pure bipartite states in the Schmidt-decomposed form as in Eq.~\eqref{Schmidt_bipartite} and using the modular Hamiltonian, we can express it as
\begin{align}
    C_{E}(\Psi_{AB}) &=\rm tr(\rho_{A}(\log\rho_{A})^{2}) - (tr(\rho_{A}\log\rho_{A}))^{2}\nonumber\\
    &=p(1-p)\big(\log\big(\frac{p}{1-p}\big)\big)^{2} .
\end{align}

We can define the rate of capacity of entanglement as
\[\frac{{\rm d}C_{E}}{{\rm d}t} =\frac{{\rm d}C_{E}}{{\rm d}p}\frac{{\rm d}p}{{\rm d}t}, \] where 
\begin{equation}
 \frac{ {\rm d} C_{E}}{ {\rm d} p} = (1-2p)\big(\log\frac{p}{1-p}\big)^{2}+2\log\frac{p}{1-p} \nonumber\\   
\end{equation}
which diverges for $p\to\{0\}\cup\{1\}$.

Let ${\Gamma_{C}}$ denote the rate of capacity of entanglement, i.e, $\Gamma_{C}:=\frac{{\rm d} C_{E}(t)}{{\rm d}t}$. From the earlier result, using the transformed Hamiltonian, we have
\begin{align}
{\Gamma_{C}} =\hspace{0.1cm} & 2\sqrt{p(1-p)}\Big((1-2p)\big(\log\frac{p}{1-p}\big)^{2} \nonumber\\
&+2\,\log\frac{p}{1-p}\Big)\,|h(H,\phi,\chi)|.
\end{align}
Thus, it will not diverge with this form for $p=0\,\text{or}\,1$.

It should be clear that local terms corresponding to \(\vec{\alpha},\vec{\beta}\) in Eq.~\eqref{bipartite_Hamiltonian}  give no contribution to $h_{max}$ with the given Schmidt-decomposed form of the bipartite state. Trying to determine $h_{max}$ in terms of \(\mu_{1,2,3}\), we get
\begin{equation}
    h(H,\phi,\chi)=\sum_{k=1}^{3}\mu_{k}\bra{\phi}\sigma_{k}^{A}\ket{\phi^{\perp}}\bra{\chi}\sigma_{k}^{B}\ket{\chi^{\perp}}.
\end{equation}
The maximum is reached for when \(\ket{\chi}=\ket{\phi^{\perp}}\). Further utilizing completeness condition \(\ket{\phi}\bra{\phi^{\perp}}+\ket{\chi}\bra{\chi^{\perp}}=I\), we get the expression
\begin{equation}
    h(H,\phi)=\sum_{k=1}^{3}\mu_{k}-\sum_{k=1}^{3}\mu_k \bra{\phi}\sigma_{k}\ket{\phi}^{2}.
\end{equation}
It can be further inferred from \(\mu_1 \geq\mu_2 \geq\mu_3\) that maximum value is reached for when \(\ket{\phi}=\ket{0}\) or \(\ket{1}\), which gives us
\begin{equation}
    h_{max} = \mu_1 + \mu_2 .
\end{equation}
Thus, the state that provides maximum rate of capacity of entanglement and the corresponding rate are given by
\begin{equation}
    \ket{\Psi_{E}} = \sqrt{p}\ket{01}+i\sqrt{1-p}\ket{10} ,
\end{equation}
\begin{align}
     {\Gamma_{C}}_{max} =\hspace{0.1cm} &\frac{{\rm d} C_{E}}{{\rm d} t}\Big{|}_{max} \nonumber\\ 
     =\hspace{0.1cm} & 2(\mu_1 +\mu_2)\sqrt{p(1-p)} \nonumber\\
     & \Big[(1-2p)\big(\log\frac{p}{1-p}\big)^{2} + 2\log\frac{p}{1-p}\Big] .
\end{align}
The maximum rate ${\Gamma_{C}}_{max}$ is obtained here for $p_0 \simeq 0.0045$ which maximizes $f(p)$ to $f(p_0 )\simeq 1.2108$ for the corresponding $\ket{\Psi_{max}}$. The capacity of entanglement for this maximum rate is ${C_E}(p_0 ) \simeq 0.1306$.\\

It has been shown that if we can allow local operations which can entangle each qubit with local ancilla, that can increase
the $\Gamma_{max}$ for certain kinds of Hamiltonian \cite{rate_ent}. We shall begin by generalizing the formulas for multilevel systems which contains the ancillas and the qubits. Consider a state $\ket{\Psi}_{AB}$ with the Schmidt-decomposition \(\ket{\Psi}_{AB}=\sum_{n=1}^{N}\sqrt{\lambda_{n}}\ket{\phi_n}\ket{\chi_n}\). Again, the capacity of entanglement only depends on the Schmidt coefficients \(\lambda_n \geq 0\). Using the definition of capacity of entanglement rate in  Eq.~\eqref{rate_capacity}, we have
\begin{align}
   \Tilde{\Gamma}_{C} &= \frac{{\rm d}C_{E}}{{\rm d}t} = \sum_{n=1}^{N}\frac{\partial C_{E}}{\partial \lambda_{n}}\frac{{\rm d} \lambda_{n}}{{\rm d}t},\nonumber \\&=\frac{1}{N}\sum_{n,m=1}^{N}\Big[\frac{\partial C_{E}}{\partial \lambda_{n}}-\frac{\partial C_{E}}{\partial \lambda_{m}}\Big]\frac{{\rm d} \lambda_{n}}{{\rm d} t}.
   \label{gamma_tilda}
\end{align}
Using the Schr\"{o}dinger equation, we find
\begin{equation}
\frac{{\rm d} \lambda_{n}}{{\rm d} t}=2\sum_{m=1}^{N}\sqrt{\lambda_{n}\lambda_{m}}\hspace{.1cm}{\rm Im}\big[\bra{\phi_{n},\chi_{n}}H\ket{\phi_{m},\chi_{m}}\big].
\end{equation}
Now, let us consider one such example where adding ancillas allows one to increase capacity of entanglement more efficiently. Let us consider the case in which the ancillas are also qubits. Letting \(\lambda_{1}=p\) and \(\lambda_2 =\lambda_3 =\lambda_4 =(1-p)/3\), Eq.~\eqref{gamma_tilda} simplifies to
\begin{equation}
    \Tilde{\Gamma}=\Tilde{f}(p)\Tilde{h}(H,\phi_n ,\chi_n ),
\end{equation} where
\begin{equation}
    \Tilde{f}(p) = 2\sqrt{p(1-p)/3}\bigg[(1-2p)\log^{2}{\frac{3p}{1-p}}+2\log{\frac{3p}{1-p}} \bigg],
\end{equation}
\begin{equation}
    \Tilde{h}(H,\phi_n ,\chi_n ) = \sum_{n=2}^{4}{\rm Im}[\bra{\phi_{n},\chi_{n}}H\ket{\phi_{m},\chi_{m}}].
\end{equation}
We have a freedom to choose the phase of states \(\ket{\phi_n }\) such that all terms add with the same sign thus allowing us to replace the imaginary parts of the above terms by their absolute values, i.e., \(\Tilde{f}(p)\) by \(\left|\Tilde{f}(p)\right|\). We find that \(\Tilde{p}_0 \simeq 0.6036\) corresponding to capacity of entanglement \(C_{E}(\Tilde{p}_0)\simeq 0.5523\) maximizing $\Tilde{f}(p)$ to \(\left|\Tilde{f}(p_0 )\right| \simeq 1.4459\). Further, proceeding to maximize $\Tilde{h}$, we obtain that the maximum value is \(\Tilde{h}_{max}=\mu_1 +\mu_2 +\mu_3\), which occurs when \(\ket{\phi_n }\) and \(\ket{\chi_n }\) are both orthogonal maximally entangled states between the qubit and the ancilla. 

Upon comparing the cases in which ancillas are used to those in which they are not used, we can either have \(\left|\Tilde{f}(\Tilde{p}_0)\right|\geq \left|f(p_0 )\right|\) or \(\Tilde{h}_{max}\geq h_{max}\). For the case when $\mu_3 \neq 0$, we can use ancillas to increase the maximum rate of capacity of entanglement $\Gamma_{max}$ as well as $\Gamma$ for a given capacity of entanglement of the state $\ket{\Psi}$.

\section{Bound on rate of entanglement}
In this section, we will show that the capacity of entanglement plays an important role in providing an upper bound for the entanglement rate for the non-local Hamiltonian. Specifically, we will show that the entanglement rate is upper bounded by the speed of transportation of the bipartite state and the time average of square root of the capacity of entanglement. Also, this sets a quantum speed limit on the entanglement production and degradation for pure bipartite states. Thus, the capacity of entanglement has a physical meaning in deciding how much time a bipartite states takes to produce a certain amount of entanglement.

Let us consider a bipartite system initially in a pure state. Let $\ket{\Psi(0)}_{AB}$ denote the initial state of the system. We consider the dynamics generated by a non-local Hamiltonial $H_{AB}$. The time evolved state at later time $t$ is given by 
$\ket{\Psi(t)}_{AB}=U_{AB}(t)|\Psi(0)\rangle_{AB}$, where $ U_{AB}(t)=e^{-iH_{AB}t}$ with $\hbar = 1$.

Now, we apply the Heisenberg-Robertson uncertainty relation~\cite{robertson_uncertain} for two non-commuting operators \(K_{A}\) and \(H_{AB}\). This leads to
\begin{equation}
 \frac{1}{2}\left|\langle\Psi(t)|[K_{A}\otimes I_{B},H_{AB}]|\Psi(t)\rangle\right|\leq \Delta K_{A}\Delta H_{AB}\, .
\label{equ:uncertainty_relation_for_modularhamiltonian_and_non-local_Hamiltonian}
\end{equation}
Recall that the evolution of average of any self adjoint operator $O$ is given by 
\begin{equation}
    i \hbar\frac{{{\rm d} \langle O\rangle}}{{\rm d}t} = \langle[O,H]\rangle \, . \label{equ:equation_of_motion_of_K}
\end{equation}
Using Eq.~\eqref{equ:equation_of_motion_of_K} (for $O = K_{A}$) in Eq.~\eqref{equ:uncertainty_relation_for_modularhamiltonian_and_non-local_Hamiltonian}, we then obtain
\begin{equation}
    \frac{\hbar}{2}\left|\frac{{\rm d} \langle K_{A}\rangle}{{\rm d}t}\right|\leq \Delta K_{A}\Delta H_{AB}\,.
\end{equation}

Let $\Gamma(t)$ denote the rate of entanglement. Recall that the average of the modular Hamiltonian is the entanglement entropy $S_{EE}$. In terms of the entanglement rate $\Gamma(t)$, the above equation can be written as
 \begin{equation}
    \left| \Gamma(t)\right| \leq \frac{2}{\hbar}\Delta K_{A}\Delta H_{AB} \, .
 \end{equation}
Since the square of the standard deviation of modular Hamiltonian is the capacity of entanglement, so in terms of the capacity of entanglement, we can write above bound  as
\begin{equation}
    \left|\Gamma(t)\right| \leq \frac{2}{\hbar}\sqrt{C_{E}(t)} \Delta H_{AB}  \, .     
    \label{equ:entanglement_speed}
\end{equation}
To interpret the above equation, first note that $\frac{2}{\hbar} \Delta H_{AB}$ is nothing but the speed of transportation of the bipartite pure
entangled state on the projective Hilbert Space of the composite system. If we use the Fubini-Study metric for two nearby states \cite{Anandan,pati-91,pati-95}, then the infinitesimal distance between two nearby states is defined as
\begin{equation}
{\rm d}S^2 = 4\left( 1 -\left|\langle{\Psi(t)}\ket{\Psi(t+{\rm dt})}\right|^2 \right) = \frac{4}{\hbar^2} \Delta H_{AB}^2 {\rm  d}t^2.
\end{equation}

Therefore, the speed of transportation as measured by the Fubini-Study metric is given by $ V= \frac{{\rm d}S}{{\rm d}t} = \frac{2}{\hbar} \Delta H_{AB}$. Thus, the entanglement rate is upper bounded by the speed of quantum evolution \cite{Deffner_2017} and the square root of the capacity of entanglement, i.e., $|\Gamma(t)|\leq\sqrt{C_{E}(t)}V$.

It was shown in Ref~\cite{Bravyi2007} that for ancilla unassisted case, the entanglement rate is upper bounded by $c\|H\|\log d$, where $d={\rm min(dim}{\cal{H}}_{A},{\rm dim}{\cal{H}}_{B})$, $c$ is constant between 0 and 1, and $\|H\|$ is operator norm of Hamiltonian which corresponds to $p=\infty$ of the Schatten p-norm of $H$ which is defined as $\|H \|_p = [\tr \left(\sqrt{H^\dag H}\right)^p ]^\frac{1}{p}$. Now, using the fact that the maximum value of capacity of entanglement  is proportional to $S_{max}(\rho_{A})^{2}$~\cite{capacity_def}, where $S_{max}(\rho_{A})$ is maximum value of von Neuamnn entropy of subsystem which is upper bounded by $\log{d}_{A}$, where $d_{A}$ is the dimension of Hilbert space of subsystem $A$, and $\Delta H \leq \|H\|$, a similar bound on the entanglement rate can be obtained from Eq.~\eqref{equ:entanglement_speed}. Thus, the bound on the entanglement rate given in Eq.~\eqref{equ:entanglement_speed} is stronger than the
previously known bounds.

The bound on the entanglement rate can be used to provide a quantum speed limit for the creation or degradation of entanglement. The notion of quantum speed limit (QSL) decides how fast a quantum state evolves in time from an initial state to a final state \cite{Pfeifer}. Even though it was discovered by Mandelstam and Tamm \cite{Mandelstam1945}, over last one decade, there have been active explorations on generalising the notion of quantum speed limit for mixed states \cite{Wu_2018,Mondal_2016} and on resources that a quantum system might posses~\cite{RSL_2020}. Recently, the notion of generalised quantum speed limit has been defined in Ref.~\cite{dimpi-22}. In addition, the quantum speed limit for observables has been defined and it was shown that the QSL for state evolution is a special case of the QSL for observable~\cite{brij-21}. For a quantum system evolving under a given dynamics, there exists fundamental limitations on the speed for entropy \(S(\rho)\), maximal information \(I(\rho)\), and quantum coherence \(C(\rho)\) \cite{mohan-22} as well as on other quantum correlations like entanglement, quantum mutual information and Bell-CHSH correlation~\cite{vivdiv-22}. Below, we provide a speed limit bound for the entanglement entropy which can be applied for scenario where entanglement can be  generated or degraded, based on the capacity of entanglement. Our bound highlights the non-trivial role played by the capacity of entanglement in deciding the QSL.

The speed limit for entanglement entropy can be calculated from Eq.~\eqref{equ:entanglement_speed} by taking the absolute value on both the sides and integrating  over time. Thus, we have 
\begin{equation}
  \int_{0}^{T} \left | \frac{{\rm d} S_{EE}(t)}{{\rm d}t}\right|{{\rm d}t} \leq \int_{0}^{T}\frac{2}{\hbar} \sqrt{C_{E}(t)} \Delta H {\rm d}t. \label{equ:integral_over_t}\\
\end{equation}
For the time independent Hamiltonian, we obtain the following bound for the quantum speed limit for entanglement 
 \begin{align}
T \ge T_{\rm QSL}^{E} := \frac{\left|S_{EE}(T)-S_{EE}(0)\right|}{\frac{2}{\hbar}\Delta H\frac{1}{T} \int_{0}^{T} \sqrt{C_{E}(t)} {\rm d}t} .
 \label{speeed_limit_entanglement}
 \end{align}
In the case of time dependent Hamiltonian $H(t)$, we can apply the Cauchy-Schwarz inequality in Eq.~\eqref{equ:integral_over_t} and obtain the following inequality
 \begin{equation}
  \int_{0}^{T} \left | \frac{{\rm d} S_{EE}(t)}{{\rm d}t}\right|{{\rm d}t} \leq\sqrt{\int_{0}^{T}\frac{2}{\hbar}\sqrt{C_{E}(t)}{\rm dt}} \sqrt{\int_{0}^{T}\frac{2}{\hbar}\Delta H_{t} {\rm dt}} .  \\
\end{equation}
From the above inequality, we get the bound for the speed limit for entanglement entropy change as given by
 \begin{align}
T  \geq  T_{\rm QSL}^E := \frac{\left|S_{EE}(T)-S_{EE}(0)\right|}{\frac{2}{\hbar}\Delta \Bar{H}\sqrt{\frac{1}{T}\int_{0}^{T}\sqrt{C_{E}(t)}\rm dt}} \,, 
 \label{speeedlimitentanglement}
 \end{align}
 where $\Delta \Bar{H}=\frac{1}{T}\int_{0}^{T}\sqrt{\langle H(t)^2\rangle-\langle H(t)\rangle^2} \,\rm dt$, is the time averaged fluctuation in the Hamiltonian. In both these bounds (time dependent and time independent Hamiltonian) it is clear that evolution speed for entanglement generation (or degradation) is a function of capacity of entanglement $C_{E}$. Thus, we can say that $C_{E}$ controls how much time a system may take to produce certain amount of entanglement. \\
 
\begin{figure}[h!]
    \centering
    \includegraphics[width=9cm]{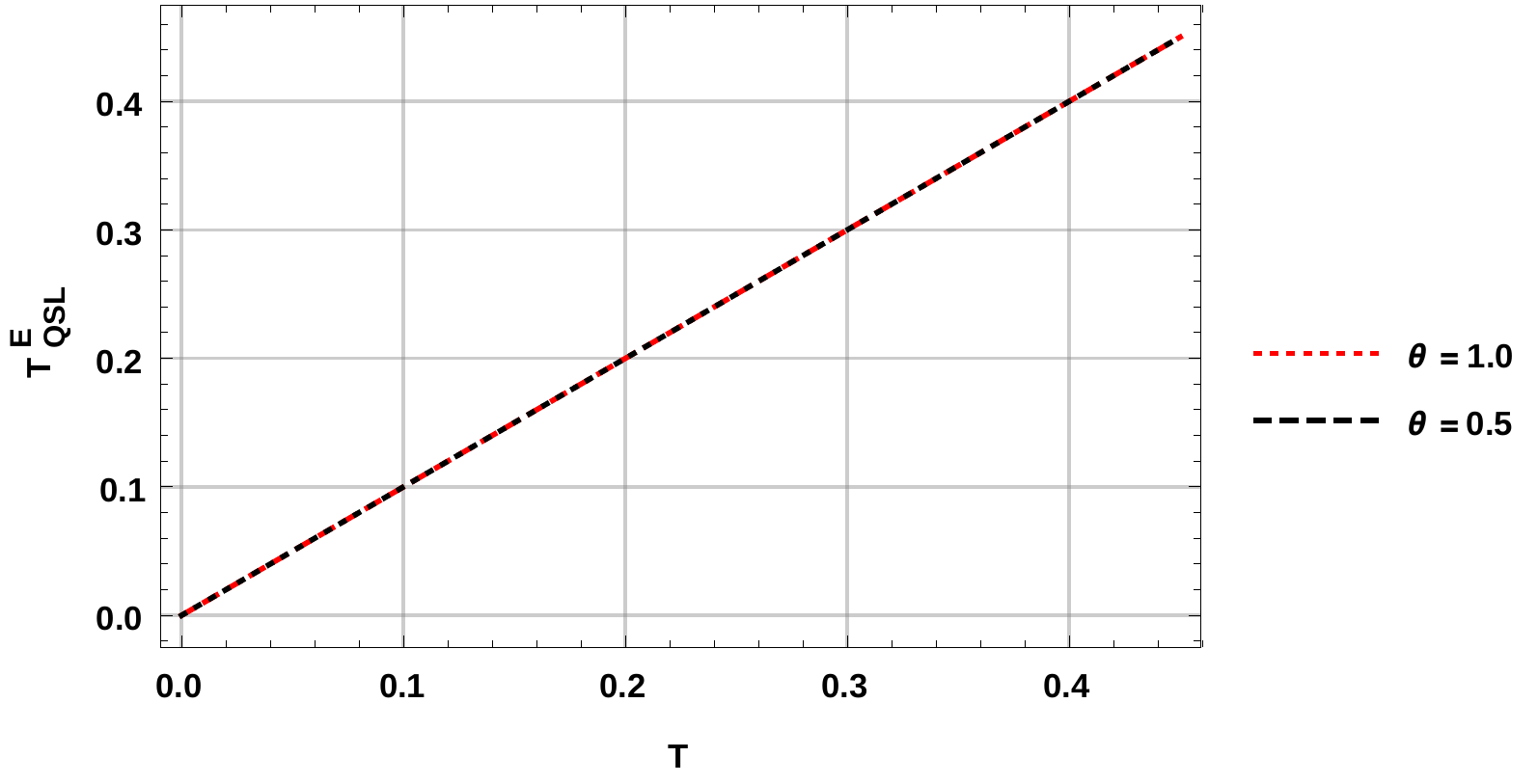}
    \caption{Here we depict $T_{QSL}^{E}$ vs \(T\) with \(p=1\) for $\theta = 0.5$ and $1.0$, which shows that our speed limit bound is tight. }
    \label{fig:speed_limit_for_non-local_hamiltonians}
\end{figure}
Now, one may ask how tight is the QSL bound for the entanglement generation of degradation? 
Here, we illustrate with a specific example that the quantum speed limit for the creation of entanglement is actually tight. Consider the initial state at $t=0$ as given in Eq.\eqref{equ:time zero state}. The time evolution of the state is given by Eq.\eqref{equ:time evolved state}. Estimating the speed limit bound on the entanglement entropy in Eq.~\eqref{speeed_limit_entanglement}, for the considered state, would need following quantities: 
\begin{equation}
    C_{E}(t) = \frac{\left(1-\eta(t)^{2}\right)\tanh^{-1}\left(\eta(t)\right)^{2}}{\ln^{2}(2)},
\end{equation}
where $\eta(t)=(1-2 p) \cos (2 \theta t)$.
\begin{align}
    \Delta H = & \theta(1-2p),\\
    S_{EE}  = & -\frac{\log_{2} \left(\left(p-\frac{1}{2}\right) \cos (2 \theta t)+\frac{1}{2}\right)}{2}\nonumber\\ 
    & +\frac{\log_{2} \left((\frac{1}{2}-p) \cos (2 \theta t)+\frac{1}{2}\right)}{2}\nonumber \\
    & +\frac{(1-2 p) \cos (2 \theta t) \tanh ^{-1}((1-2 p) \cos (2 \theta t))}{\ln (2)}.
\end{align}

The plot in Fig.~\ref{fig:speed_limit_for_non-local_hamiltonians} for $T_{QSL}^{E}$ vs $T \in [0,0.45]$ is shown under unitary dynamics through a general two qubit non-local Hamiltonian $H^{+}_{AB}$, beginning with an initial state of the system $\ket{\Psi(0)}=|0\rangle|0\rangle$ (taking $p=1$ in Eq.~\eqref{equ:time evolved state}). Our example shows that for the case of $\theta=\mu_{1}-\mu_{2}=0.5$ and $1.0$, the QSL for the entanglement creation is indeed tight and attainable.

\section{Capacity of entanglement for self inverse Hamiltonian}
In this section, we will explore the dynamics of capacity of entanglement for the self inverse Hamiltonian. Such Hamiltonians are simpler to handle and provide many interesting insights. The rate of capacity of entanglement for the self inverse Hamiltonian has been addressed. It was found that the inclusion of ancilla system lead to the enhancement of  the entanglement capability in Ref.\cite{rate_ent}, but for the  Ising Hamiltonian \(H_{ising}=\sigma_z\otimes\sigma_z\) it was shown that entanglement capability is  ancilla-independent \cite{child}. This independence on ancillas of entanglement capabilities turns out to be a consequence of the self-inverse property of the Hamiltonian \(H_{ising}=H_{ising}^{-1}\). This result was generalized  to all Hamiltonian evolutions of the kind \cite{H_inv}
\begin{equation}
    H_{AB}=X_{A}\otimes X_{B}
    \label{selfinverse}
\end{equation}
such that $X_{i}=X_{i}^{-1}\in\,\mathcal{H}_{i}$ for \(i\,\in\,\{A,B\}\).
As a consequence of self-inverse property of the Hamiltonian, we have the time evolution operator ($\hbar=1$)
\begin{equation}
    U(t) = e^{-i Ht}= \cos{t}\,\mathcal{I}_{A}\otimes\mathcal{I}_{B}-i\sin{t}\,X_{A}\otimes X_{B}.
\end{equation}

Let $\ket{\Psi(0)}_{AB} $ be the initial state of the bipartite system \(AB\), which can be expressed in the Schmidt decomposition as follows
\begin{equation}
   \ket{\Psi(0)}_{AB}=\sum_{n}\sqrt{\lambda_{n}}\ket{\psi_{n}}_{A}\otimes\ket{\phi_{n}}_{B}. \label{equ:Schmidt_Decomposition_of_state_in_Self_Inverse_Hamiltonian_case}
\end{equation}  
Let $\rho_{AB}(t)$ denote the density operator at time $t$. The time evolution of $\rho_{AB}(t)$ is governed by the Liouville-von Neumann equation given as
\begin{equation}
    \frac{{\rm d}\rho_{AB}(t)}{{\rm d}t}= -i [H_{AB},\rho_{AB}(t)],
\end{equation}
where $H_{AB}$ is the non-local Hamiltonian of the composite system. The dynamics of the reduce density operator $\rho_{B}$ (or $\rho_{A}$) can be obtained from above equation by tracing out $A~(\text{or}~B)$, which is given by
 \begin{equation}
     \frac{{\rm d}\rho_{B}}{{\rm d}t}=-i\tr_{A}[H_{AB},\rho_{AB}(t)] .
     \label{equ:Liouville_von-Neumann_equation}
 \end{equation}
Now, first we will calculate an upper bound on rate of capacity of entanglement for unitary evolution and then we will address the case of self inverse Hamiltonian. To calculate an upper bound on $C_{E}$, first we differentiate both the sides of Eq.~\eqref{equ:Entanglement_Capacity} with respect to time, this leads to
\begin{align}
\frac{{\rm d} C_{E}(t)}{{\rm d}t}&= \frac{\rm d}{{\rm d}t}\left(\langle K_{A}^{2}\rangle-\langle K_{A}\rangle^{2}\right)\nonumber\\
&=\frac{\rm d}{{\rm d}t}\left(\tr(\rho_{A}(-\log\rho_A )^{2})\right)-\frac{\rm d}{{\rm d}t}\left(-\tr(\rho_{A}\log\rho_{A})\right)^{2} \nonumber \\
&=\frac{\rm d}{{\rm d}t}\left(\tr(\rho_{A}(-\log\rho_A )^{2})\right)- 2S(\rho_{A})\frac{\rm d}{{\rm d}t}S(\rho_{A})\nonumber\\
&=\frac{\rm d}{{\rm d}t}\left(\tr(\rho_{A}(-\log\rho_A )^{2})\right)- 2S(\rho_{A})\Gamma(t) \,,
\label{equ:Capacity_Derivative}
\end{align}
where $\Gamma(t)$ is the rate of entanglement. Now, we use the fact that logarithm of an operator $\rho$ can be represented by
\begin{align}
    \log \rho &= \int_{0}^{\infty} {{\rm d}s} \left(\frac{1}{(s+1)\cal{I}}-\frac{1}{(s\cal{I}+\rho)}\right), \label{equ:integral_representation_of_log_rho}
\end{align}
where $\cal{I}$ is the identity operator. We use the above equation to compute the first terms on the right hand side of Eq.~\eqref{equ:Capacity_Derivative}. This can be expressed as

\begin{widetext}
\begin{align}
    \frac{\rm d}{{\rm d}t}\left(\tr(\rho_{A}(-\log\rho_A )^{2})\right)&=\tr\left(\frac{\rm d}{{\rm d}t}\left(\rho_{A}(-\log\rho_A )^{2}\right)\right) ,\nonumber\\
    &=\tr\left(\left(\Dot{\rho}_{A}(\log\rho_A )^{2}\right)+\rho_{A}\frac{\rm d}{{\rm d}t}(\log\rho_{A})^{2}\right),\nonumber\\
    &=\tr\left(\left(\Dot{\rho}_{A}(\log\rho_A )^{2}\right)\right)
    +\tr\left(\rho_{A}\frac{\rm d}{{\rm d}t}\left( \int_{0}^{\infty} {{\rm d}s} \left(\frac{1}{(s+1)\cal{I}}-\frac{1}{(s\cal{I}+\rho_{A})}\right)\right)(\log\rho_{A})\right)
    \nonumber\\
    &\hspace{0.35cm}+\tr\left(\rho_{A}(\log \rho_{A})\frac{\rm d}{{\rm d}t}\left( \int_{0}^{\infty} {{\rm d}s} \left(\frac{1}{(s+1)\cal{I}}-\frac{1}{(s\cal{I}+\rho_{A})}\right)\right)\right),\nonumber\\
    &=\tr\left(\left(\Dot{\rho}_{A}(\log\rho_A )^{2}\right)\right)
    +\tr\left(\rho_{A}\left( \int_{0}^{\infty} {{\rm d}s} \left(\frac{1}{(s\mathcal{I}+\rho_{A})}\Dot{\rho}_{A}\frac{1}{(s \mathcal{I}+\rho_{A})}\right)\right)(\log\rho_{A})\right)
    \nonumber\\
    &\hspace{0.35cm}+\tr\left(\rho_{A}(\log \rho_{A})\left( \int_{0}^{\infty} {{\rm d}s} \left(\frac{1}{(s \mathcal{I}+\rho_{A})} \Dot{\rho}_{A}\frac{1}{(s \mathcal{I}+\rho_{A})}\right)\right)\right),\nonumber\\
    &=\tr\left(\left(\Dot{\rho}_{A}(\log\rho_A )^{2}\right)\right)+2\tr\left(\Dot{\rho_{A}}\log\rho_{A}\right).
\end{align}
\end{widetext}

The second term on the right hand side of above equation is the rate of the entropy~\cite{Das_2018}, so we rewrite Eq.~\eqref{equ:Capacity_Derivative} as 
\begin{align}
\frac{{\rm d} C_{E}(t)}{{{\rm d}}t}&
=\tr(\Dot{\rho_{A}}(-\log\rho_{A})^2)+2\tr(\Dot{\rho_A}
\log\rho_A)(1+S(\rho_A)).
\end{align}
Now, we consider the case where $\rho$ is full rank, then the first term of above equation can be simplified as 
\begin{eqnarray}
      \tr[\Dot{\rho}(\log \rho)^2] 
      =&\sum_{i}\bra{i}\Dot{\rho} \ket{i} (\log \lambda_i)^2 \nonumber\\
      \leq& k_{max}^{2} \sum_{i} \bra{i}\Dot{\rho}\ket{i} \nonumber\\
      =& k_{max}^{2} \tr[\Dot{\rho}_A]=0,
\end{eqnarray}
where $k_{max}$ is the maximum of the eigenvalues of the modular Hamiltonian. We then obtain an upper bound on the capacity of entanglement as 
\begin{align}
\left|{\Gamma_{C}}\right|& \leq\left|2\tr(\Dot{\rho_A}
\log\rho_A)(1+\log d_{A})\right|\nonumber\\
&=\left|2\Gamma(t)(1+\log d_{A})\right|.\label{equ:capaciti_rate}
\end{align}
Using Eq.\eqref{equ:entanglement_speed} , we can give an upper bound on the rate of capacity of entanglement as given by 
\begin{align}
    \left|{\Gamma_{C}}\right|&\leq 2\sqrt{C_{E}}V(1+\log d_{A}),
\end{align}
where $V = \frac{2}{\hbar}\Delta H_{AB}$ is the speed of bipartite quantum state.

For the ancilla unassisted case, the entanglement rate $\Gamma(t)$ is upper bounded by $c||H||\log d$~(see Ref.~\cite{Bravyi2007}). Then, the upper bound on the rate of capacity of entanglement $\Gamma_{C}$ becomes
\begin{equation}
    \left|\Gamma_{C}\right|\leq 2c||H||\log d(1+\log d_{A}).
\end{equation}
Now we will find the upper bound on $\Gamma_{C}$ for self inverse Hamiltonians. The maximum entanglement rate $\Gamma(t)$ for the self inverse Hamiltonian $H=X_{A}\otimes X_{B}$ was calculated in Ref.~\cite{H_inv}. It is given by $\Gamma(t) \le \beta$, where 
\begin{equation}
    \beta = 2 \max_{x\in [0,1]} \sqrt{x(1-x)}\log\frac{x}{1-x}. \\
\end{equation}
Therefore, the bound on $\Gamma_{C}$ can be expressed as
\begin{equation}
    \left|\Gamma_{C}\right|\leq 2\beta(1+\log d).
\end{equation}
 This bound is independent of the details of the initial state but uses the self-inverse nature of the non-local Hamiltonian.
 
 \section{Capacity of entanglement for mixed states}
In the previous section, we used definition of $C_{E}$ for pure states. Here, we generalise the definition for the case of mixed states in such a way that it reduces to the previous definition for pure states. For this, we use the relative entropy of entanglement since it reduces to the entanglement entropy for pure states. The relative entropy of entanglement is defined in Ref.~\cite{entropy} and further expanded for arbitrary dimensions in Ref.~\cite{relent_alld}. This is given by
\begin{equation}
    E_{R}(\rho_{AB})= \min_{\sigma_{AB} \in {\rm SEP}} {S(\rho ||\sigma)},
\end{equation}
where \(\rm SEP\) is set of all separable or positive partial transpose (PPT) states and $S(\rho ||\sigma) = {\rm Tr}(\rho\log\rho -\rho\log\sigma).$ 
Operationally, the relative entropy of entanglement quantifies the extent to which a given mixed entangled state can be
distinguished from the closest state which is either separable or has a positive partial transpose (PPT). 
Also, this is an entanglement monotone and it is asymptotically continuous.

In the following, we shall denote the state in $\rm SEP$ for which the the minimum is attained for a given $\rho_{AB}$ as $\rho_{AB}^*$. Then, we can write $E_{R}(\rho_{AB})$ as 
\begin{equation}
    E_{R}(\rho_{AB}) = \min_{\sigma_{AB}\in \rm SEP} {S(\rho_{AB}||\sigma_{AB})}=S(\rho_{AB}||\rho_{AB}^{*}).
\end{equation}

Now, we claim that the capacity of entanglement for mixed states is given by
\begin{equation}
\begin{split}
     C_{E}(\rho_{AB}) & = \tr(\rho_{AB}(\log{\rho_{AB}}-\log{\rho^{*}_{AB}})^{2}) \label{mixed} \\
   &\hspace{0.35cm} -\tr(\rho_{AB}(\log{\rho_{AB}}-\log{\rho^{*}_{AB}}))^{2}.
\end{split}
\end{equation}

We will now show that this agrees with the definition of capacity of entanglement for pure states. The relative entropy of entanglement is given by
\begin{eqnarray}
      E_{R}(\rho_{AB}) &=& \tr(\rho_{AB}(\log{\rho_{AB}}-\log{\rho_{AB}^{*}}) \nonumber \\
             &=& \langle \log{\rho_{AB}}-\log{\rho_{AB}^{*}} \rangle.
\end{eqnarray}

For a pure state, the density operator $\rho_{AB}$ is given by
\begin{equation}
\rho_{AB} =\ket{\Psi}_{AB}\bra{\Psi}=\sum_{i,j}\sqrt{p_{i}p_j}|\phi_i\rangle\langle\phi_j|\otimes|\chi_i\rangle \langle \chi_j|_{AB}.
\end{equation}
The expression for $\rho^{*}_{AB}$ for $\rho_{AB}$ is known \cite{vedral_puri} and given as follows
\begin{align}
 \rho_{AB}^{*}  = \sum_{k} p_{k} |\phi_{k}\rangle\langle\phi_{k}|  \otimes|\chi_{k}\rangle\langle\chi_{k}|_{AB}.
\end{align}
The first term of Eq.~\eqref{mixed} is given by
\begin{align}
   \langle(\log{\rho_{AB}} & -\log{\rho_{AB}^{*}})^2\rangle \nonumber \\ 
   =\hspace{0.1cm} & {_{AB}}\langle \Psi|\Big[\big(\log{|\Psi\rangle_{AB}\langle\Psi|})^{2}+(\log{\rho_{AB}^{*}})^{2}\big) \hspace{0.1cm} \nonumber\\ 
   & -\big(\log{|\Psi\rangle_{AB}\langle\Psi|}\log{\rho_{AB}^{*}}\nonumber \\
   & +\log{\rho_{AB}^{*}}\log{|\Psi\rangle_{AB}\langle\Psi|}\big)\Big]|\Psi\rangle_{AB} .
   \label{first}
\end{align}
Defining $A_{\Psi}=|\Psi\rangle_{AB}\langle\Psi| - \mathcal{I}$, we get
\begin{align*}
 {_{AB}}\langle \Psi|(\log |\Psi \rangle_{AB}\langle \Psi |)= {_{AB}}\langle \Psi| \Big[A_{\Psi} -\frac{(A_{\Psi})^2}{2}+...\Big]=0.
\end{align*}
This leads to
\begin{equation}
    {_{AB}}\langle \Psi|(\log |\Psi \rangle_{AB}\langle \Psi |)^2 |\Psi \rangle_{AB} =0.
\end{equation}
with the only surviving term in Eq.~\eqref{first} is $\langle\Psi|(\log{\rho_{AB}^{*}})^{2}|\Psi\rangle_{AB}$. 

Now, we have
\begin{align*}
(\log{\rho_{AB}^{*}})^{2}&=\sum_{k}(\log{p_{k}})^{2}|\phi_{k}\rangle_{A}\langle\phi_{k}|  \otimes|\chi_{k}\rangle_{B}\langle\chi_{k}|\\
 \langle (\log{\rho_{AB}^{*}})^2\rangle & = \sum_{i,j,k}\sqrt{p_{i}p_{j}}(\log{p_{k}})^{2}\delta_{ik}\delta_{jk},\\
&=\sum_{k}p_{k}(\log{p_k})^{2} = \langle(\log{\rho_{A}})^{2}\rangle .   
\end{align*}


The second term of Eq.~\eqref{first} is equal to $E(\rho_{AB})^2$ for pure states. Thus, for $\rho_{AB}=|\Psi\rangle\langle\Psi|_{AB}$, we have,
\begin{equation}
 C_{E}=\langle(\log{\rho_A})^{2}\rangle-\langle \log{\rho_{A}}\rangle^{2}
\end{equation}
which agrees with the expression for the capacity of entanglement for the pure bipartite states. 

It may be worth noting that the capacity of entanglement for mixed state can also be expressed as the variance of the shifted modular Hamiltonian for the joint system. Upon defining the modular Hamiltonian for the composite state $\rho_{AB}$ and \(\rho_{AB}^{*}\) as \(K_{AB}=-\log{\rho_{AB}}\) and \(K_{AB}^{*}=-\log{\rho_{AB}^{*}}\), we have 
\begin{align}
C_{E} & = \tr[\rho_{AB} (K_{AB}-K_{AB}^{*})^{2} ]  - \tr[\rho_{AB} (K_{AB}-K_{AB}^{*})]^2 \nonumber \\ & = \langle(K_{AB}-K_{AB}^{*})^{2}\rangle - \langle K_{AB}-K_{AB}^{*}\rangle^{2} \nonumber \\
& = \langle\tilde{K}_{AB}^{2}\rangle - \langle\tilde{K}_{AB}\rangle^{2},
\end{align}
where $\tilde{K}_{AB} = K_{AB}-K_{AB}^{*}$, is the shifted modular Hamiltonian for the composite system. This provides another meaning for the capacity of entanglement for the mixed state.

Now, we illustrate the capacity of entanglement for mixed state using the above definition. For general mixed entangled states, it is not always easy to find the closest separable state. However, for those cases where we know the closest separable state, we can compute the
capacity of entanglement. 

\begin{figure}[h!]
    \centering
    \includegraphics[width=9cm]{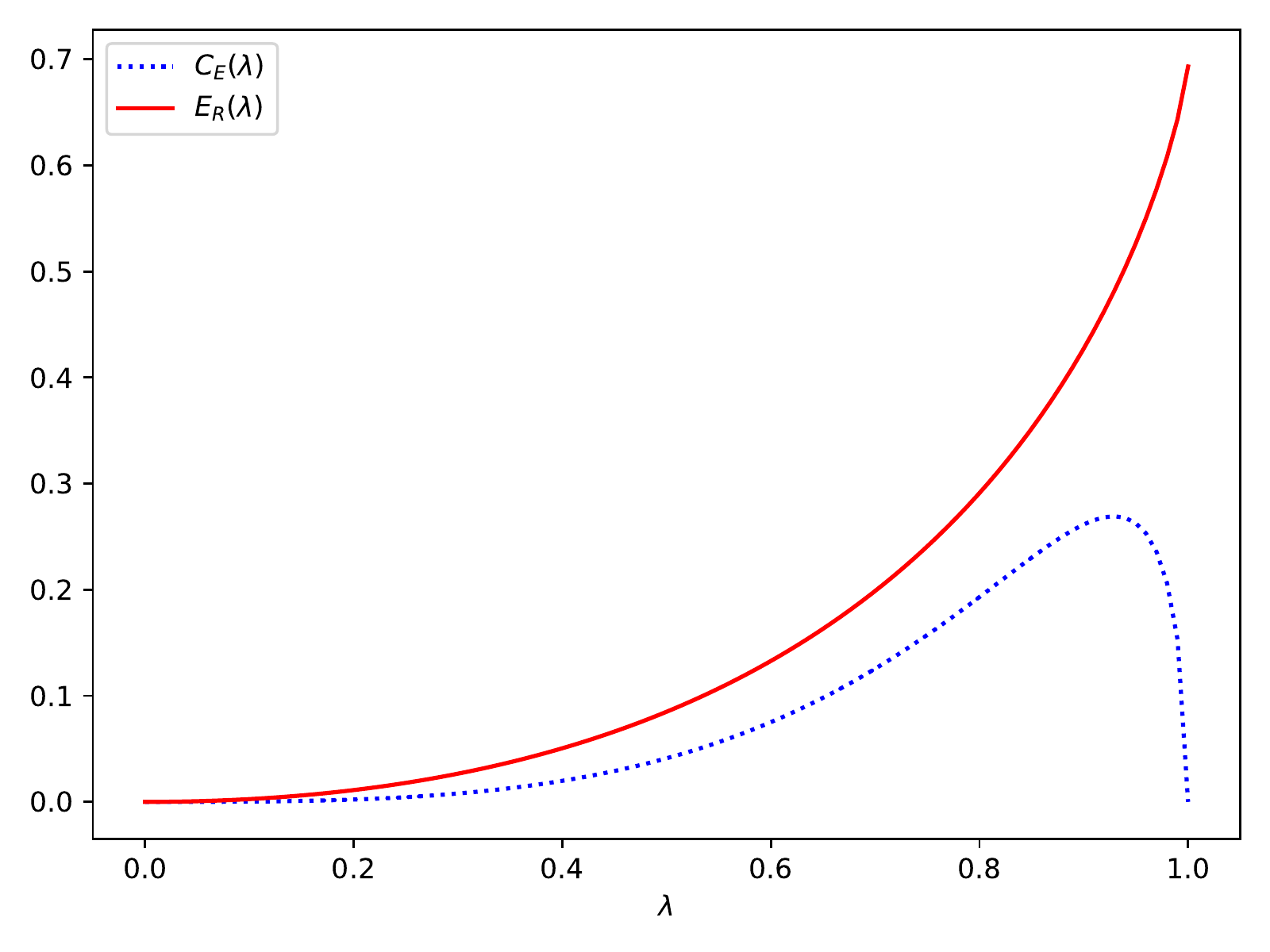}
    \caption{Plot for capacity of entanglement($C_E $) and relative entropy of entanglement ($E_{R}$) vs $\lambda \in [0,1]$ for $\rho_{AB}$ in Eq.~\eqref{equ:eg mix state}.}
    \label{fig:C_E+S}
\end{figure}
\begin{figure}[h!]
    \centering
    \includegraphics[width=9cm]{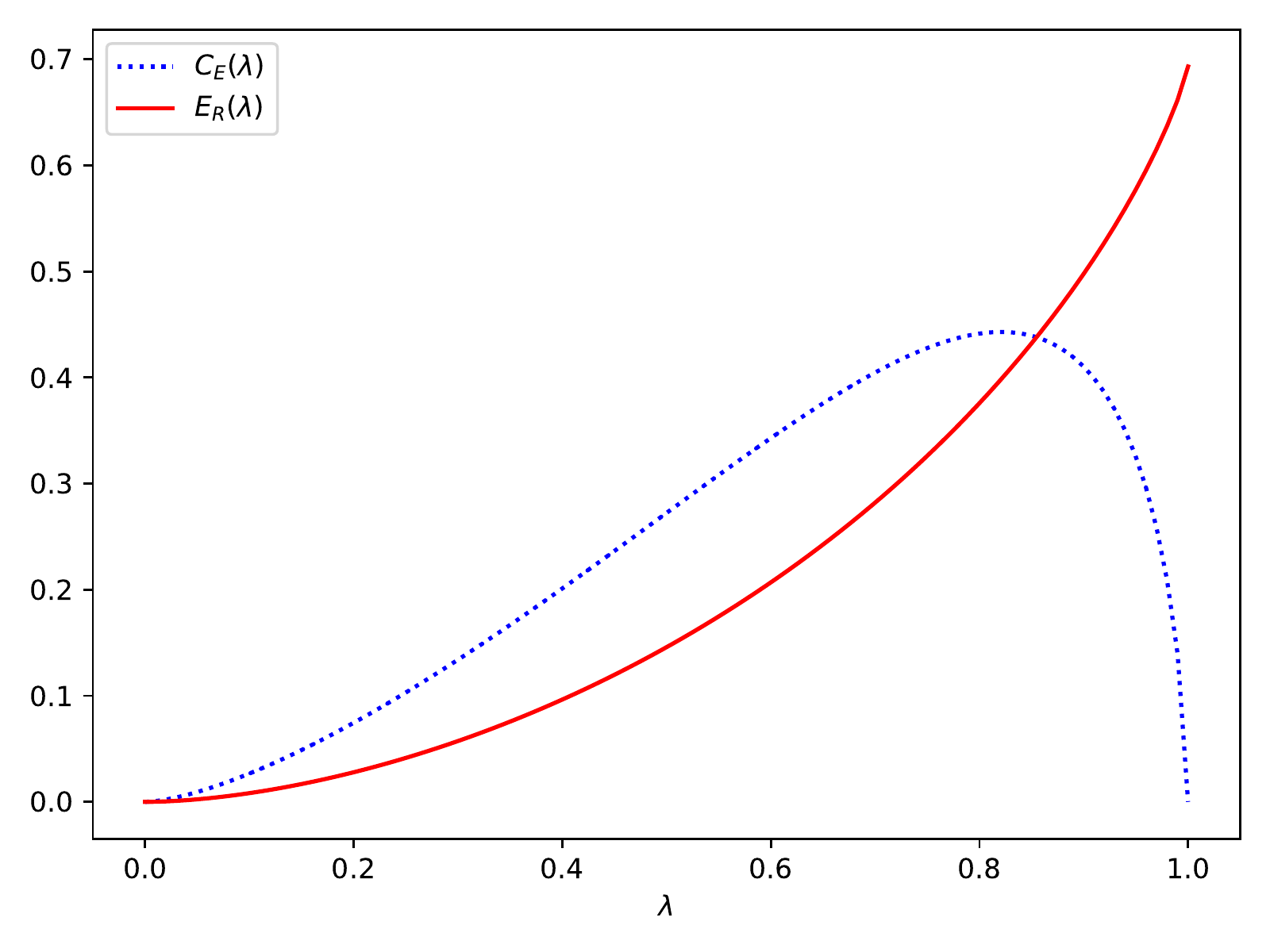}
    \caption{Plot for capacity of entanglement($C_E $) and relative entropy of entanglement~($E_{R}$) vs $\lambda \in [0,1]$ for $\rho_{AB}$ in Eq.~\eqref{equ:eg mix state2}.}
    \label{fig:C_E+S_2}
\end{figure}

Let us consider a mixed entangled state as given by
\begin{align}
    \rho_{AB} =\hspace{0.1cm} &\lambda \ket{\phi^{+}}\bra{\phi^{+}}+(1-\lambda)\ket{01}\bra{01}
    \label{equ:eg mix state}
\end{align}

where \(\ket{\phi^{+}}=\frac{1}{\sqrt{2}}(\ket{00}+\ket{11})\) is one of the four Bell states.
The corresponding closest separable state which minimizes quantum relative entropy with $\rho_{AB}$ \cite{vedral_puri} is given by

\begin{align}
    \rho^{*}_{AB} = &\hspace{0.2cm} \frac{\lambda}{2}\bigg(1-\frac{\lambda}{2}\bigg)\Big(\ket{00}\bra{00}+\ket{00}\bra{11}+\ket{11}\bra{00} \nonumber \\
    &+\ket{11}\bra{11}\Big)+\bigg(1-\frac{\lambda}{2}\bigg)^{2}\ket{01}\bra{01}+\frac{\lambda^{2}}{4}\ket{10}\bra{10}.
\end{align}

The expression for the relative entropy of entanglement for this example is given by
\begin{equation}
    E_{R}\left(\lambda\right) = (\lambda-2)\ln\left(1-\frac{\lambda}{2}\right)+(1-\lambda)\ln\left(1-\lambda\right).
\end{equation}

Consider another example of a mixed state
\begin{align}
    \rho_{AB} =\hspace{0.1cm} &\lambda \ket{\phi^{+}}\bra{\phi^{+}}+(1-\lambda)\ket{00}\bra{00}.
    \label{equ:eg mix state2}
\end{align}
The closest separable state minimizing relative entropy for this case is of the form \cite{vedral_puri}
\begin{equation}
    \rho_{AB}^{*} = \bigg(1-\frac{\lambda}{2}\bigg)\ket{00}\bra{00}+\frac{\lambda}{2}\ket{11}\bra{11}
\end{equation}
The relative entropy of entanglement in this case can be analytically be found and given as
\begin{align}
    E_{R}\left(\lambda\right) =\hspace{0.1cm}& s_{+}\ln(s_{+})+s_{-}\ln(s_{-})\nonumber \\&-2\left(1-\frac{\lambda}{2}\right)\ln\left(1-\frac{\lambda}{2}\right) ,
\end{align}
where 
\[s_{\pm}=\frac{1\pm\sqrt{1-2\lambda\left(1-\frac{\lambda}{2}\right)} }{2}. \]

The detailed expression for the capacity of entanglement for $\rho_{AB}$ in Eq.~\eqref{equ:eg mix state} and Eq.~\eqref{equ:eg mix state2} are very complicated. For the purpose of illustration we have provided numerical plots for the same. From the behaviour of plots in Fig.~\ref{fig:C_E+S} and Fig.~\ref{fig:C_E+S_2}, it can be inferred that for $\lambda \in \{0,1\}$, the cases where all non-zero eigenvalues of the state are same and thus the state becomes either pure or maximally mixed, and for such flat states, the capacity of entanglement vanishes. We leave the detailed investigation for the mixed state case for future work.

\section{Conclusions} 
Undoubtedly, study of quantum entanglement for bipartite and multipartite states is one of the prime area of research over last several decades. Even though the dynamics of entanglement for non-local Hamiltonians has been addressed earlier, the question of dynamics of the capacity of entanglement has not been studied before. The notion of the capacity of entanglement is a very useful quantity and this can be regarded as the quantum information theoretic counterpart of the heat capacity. For any bipartite pure state, the capacity of entanglement is the variance of the modular Hamiltonian in the reduced state of any of the subsystem. In this paper, we have studied the dynamics
of the capacity of entanglement under non-local Hamiltonian. Our results answers a very pertinent question on the capacity of entanglement that the system can possess when it evolves in time under a non-local Hamiltonian. The capacity of entanglement has another meaning in deciding the upper bound for the entanglement rate. We have shown that the quantum speed limit for creating the entanglement is not only governed by the fluctuation in the non-local Hamiltonian, i.e., the speed of transportation of bipartite state, but also depends inversely on the time average of the square root of the capacity of entanglement. In addition, we have discussed the capacity of entanglement for self-inverse Hamiltonian and found an upper bound for this case on the rate of capacity of entanglement. We have also generalised this quantity for bipartite mixed states based on the relative entropy of entanglement, which reduces to known form for pure states case. In future, it will be
worth exploring this notion which will have useful applications in other areas of physics. 

\begin{acknowledgements}
DS, SB and VP thank Brij Mohan and Ujjwal Sen for useful discussions. AKP acknowledges support of the J.C. Bose Fellowship from the Department of Science and Technology (DST), India under Grant No.~JCB/2018/000038 (2019–2024).
\end{acknowledgements}

\bibliography{main.bib}

\end{document}